\documentclass[preprint,twocolumn]{aa}
\usepackage[varg]{txfonts}
\usepackage{natbib}
\bibpunct{(}{)}{;}{a}{}{,} 
\usepackage{url}
\usepackage{amsmath,bm}
\usepackage{color,xcolor}

\usepackage[colorlinks, linkcolor=blue, anchorcolor=blue, citecolor=blue]{hyperref}
\usepackage{breakurl}

\definecolor{darkcyan}{RGB}{0,102,204}

\begin{document}

\title{The configuration and failed eruption of a complex magnetic flux rope above a $\delta$ sunspot region} 

\author{Lijuan Liu\inst{1,2} 
     \and Jiajia Liu\inst{3}
     \and Jun Chen\inst{4}
     \and Yuming Wang\inst{4, 2}
     \and Guoqiang Wang\inst{5}
     \and Zhenjun Zhou\inst{1,2}
     \and Jun Cui\inst{1,2,6}}


\institute{Planetary Environmental and Astrobiological Research Laboratory (PEARL), School of Atmospheric Sciences, Sun Yat-sen University, Zhuhai, Guangdong, 519082, China\\
  \email{liulj8@mail.sysu.edu.cn}
  \and CAS center for Excellence in Comparative Planetology, China
  \and Astrophysics Research Centre, School of Mathematics and Physics, Queen's University, Belfast, BT71NN 
  \and CAS Key Laboratory of Geospace Environment, Department of Geophysics and Planetary Sciences, University of Science and Technology of China, Hefei, Anhui, 230026, China
  \and Institute of Space Science and Applied Technology, Harbin Institute of Technology, Shenzhen, China
  \and CAS Key Laboratory of Lunar and Deep Space Exploration, National Astronomical Observatories, Chinese Academy of Sciences, Beijing, 100012, China}

\date{Received 2 January 2021}

\abstract {} 
{We investigate the configuration of a complex flux rope 
above the $\delta$ sunspot region in NOAA AR 11515, and its eruptive expansion 
during a confined M5.3-class flare.} 
{We study the formation of the $\delta$ sunspot using the continuum intensity images and photospheric vector magnetograms provided by the Helioseismic and Magnetic Imager on board the {\it Solar Dynamics Observatory (SDO)}. We employ the extreme-ultraviolet and ultraviolet images provided by the Atmospheric Imaging Assembly (AIA) on board {\it SDO}, and the hard X-ray emission recorded by the {\it Reuven Ramaty High-Energy Solar Spectroscopic Imager} to investigate the eruptive details. 
The coronal magnetic field is extrapolated from the photospheric field using a non-linear force free field (NLFFF) method, based on which the flux rope is identified through calculating the twist number $T_w$ and squashing factor $Q$. 
We search the null point via a modified Powell hybrid method.} 
{The collision between two newly emerged spot groups form the $\delta$ sunspot. 
A bald patch (BP) configuration forms at the collision location between one umbra and the penumbra, 
above which a complex flux rope structure is identified. 
The flux rope has a multi-layer configuration, 
with one compact end and the other end 
bifurcating into different branches, 
having non-uniform $T_w$ profile which decreases 
from the core to the boundary. 
The outmost layer is 
merely sheared. 
A null point is located above the flux rope. 
The eruptive process consists of precursor flarings at a ``v''-shaped coronal structure,  
rise of the filament, and brightening below the filament, 
corresponding well with the topological structures deduced from the NLFFF, including a higher null point, a flux rope, and a BP and a hyperbolic flux tube (HFT) below the flux rope. 
Two sets of post-flare loops and three flare ribbons in the $\delta$ sunspot region further support the bifurcation configuration of the flux rope.} 
{Combining the observations and magnetic field extrapolation, we conclude that the precursor reconnection, which occurs at the null point, 
weakens the overlying confinement 
to allow the flux rope to rise, 
fitting the breakout model. 
The main phase reconnection, which may occur at the BP or HFT, 
facilitates the flux rope rising. 
The results suggest that the $\delta$ spot configuration presents an environment prone to the formation of complex magnetic configurations 
which will work together to produce activities. 
} 

\maketitle

\section{Introduction}\label{sec:intr}


Sunspots appear as dark spots on the photosphere in white light observation,  
reflecting the convection inhibition caused by the concentrations of strong magnetic flux  which constitute the solar active regions (ARs). 
They usually appear as groups, exhibiting various configurations. 
The most complex type is called $\delta$ sunspot, 
which is manifested as two umbrae of opposite polarities sharing a common penumbra~\citep{Kunzel_1960}. %
Observations reveal that regions harboring $\delta$ spots are productive in solar flares and coronal mass ejections (CMEs)~\citep{Shi_1994, Sammis_etal_2000, Takizawa_2015, Toriumi_2016, Toriumi_2019}, which are 
the two most spectacular phenomena in solar atmosphere. 
A few common features observed in $\delta$ spot regions are considered to account for their high productivity. 
For example, the opposite-signed polarities in these regions tend to be stronger and compact, forming high-gradient, strongly-sheared polarity inversion lines (PILs), 
which indeed indicate the non-potentiality of the magnetic field~\citep{Hagyard_1984, Tanaka_1991, schrijver_2007_characteristic}. 
Shearing motions along the PILs and sunspot rotations also frequently appear. They are suggested to be able to shear or twist the field~\citep{Hagyard_1984, Leka_etal_1996, Fanyh_2009, Yanxl_2015}. 
The non-potential magnetic field, 
in combination with photospheric motions, 
may create an environment apt to the magnetic reconnections and eruptions~\citep{Fangf_2015}. %

The formation conditions of $\delta$ spots are summarized into three categories by~\citet{Zirin_1987}: 
they can form when intertwined bipoles emerge at once, when satellite spots emerge into pre-existent spot regions, and when different spot groups collide. %
Numerical simulations suggest that the formation of $\delta$ spots is resulted from the emergence of twisted magnetic flux tubes from the convection zone, 
which can induce shearing motions and sunspot rotations self-consistently~\citep{Leka_etal_1996, Fangf_2015, Toriumi_2019b}. 
After formation, the coronal counterparts of $\delta$ spot regions exhibit complex configurations as well, 
sometimes are in the form of sigmoidal structures which are considered as coronal proxies of the magnetic flux rope~\citep[e.g.,][]{Joshi_2018}. The flux rope is a magnetic structure consisting of a set of helical field lines collectively wrapping around an axis with more than one full turn~\citep[][and reference therein]{Chengx_2017}.

It is suggested that the magnetic flux rope is a fundamental structure in solar eruptions~\citep[][and reference therein]{Liu_2020a}. 
For example, although flares and CMEs appear differently, 
with the former being local enhancements of electromagnetic radiation and the latter being more global expulsions of magnetized plasma, 
they are suggested to be different manifestations of a same process involving the eruption of magnetic flux ropes~\citep{Forbes_2000}.  
In the classical two-dimensional (2D) scenario of the standard flare model, the flux rope erupts to form the main body of the CME, generating current sheet at the CME wake 
through stretching the strapping field. 
The reconnection manifested as flare occurs at the current sheet~\citep{Carmichael_1964, STURROCK_1966, Hirayama_1974, Kopp_1976, Shibata_1995}. 
The standard flare model has been extended to the three-dimensional (3D) space~\citep{Aulanier_2012, Janvier_2013}, in which the current sheet where the flare occurs 
is suggested to be formed around the hyperbolic flux tube~\citep[HFT,][]{Titov_2002a}.  
The exact trigger of the eruption is explained by other models that can be roughly divided into two categories: magnetic reconnection model and ideal Magnetohydrodynamics (MHD) instability model~\cite[][and reference therein]{Chenpf_2011}. 
For the former, magnetic reconnection may occur above the sheared arcades at the null point as in the breakout model~\citep{Antiochos_1999}, or below the sheared arcades as in the tether-cutting model~\citep{Moore_2001}. In these scenarios, the flux rope forms on the fly. 
On the contrast, pre-existent flux rope is usually required in ideal MHD instability models, such as kink instability and torus instability models~\citep{Torok_2004, Kliem_2006}. 
The above models are supported by observations~\citep[][and reference therein]{Chengx_2017}. 

Observationally, 
we are not able to ``see'' the coronal flux rope directly since there is no direct measurement of the coronal magnetic field at present.  
A few observable entities such as sigmoids~\citep{Canfield_etal_1999}, filaments~\citep{Martin_1998} and hot channels~\citep{Zhangj_2012}, 
are considered as proxies of the flux ropes~\citep[][and reference therein]{Liu_2020a}. 
In situ observations of interplanetary magnetic clouds 
also give hint to the configuration of the flux ropes~\citep{Hu_2014, Wang_2018b}. 
On the other hand, 
various methods are developed to reconstruct the 3D 
coronal magnetic field~\citep[][and references therein]{Guo_2017, Wiegelmann_2017}, 
based on which one can locate and analyze the flux rope~\citep{rliu_2016}. 
It is found that the flux rope in reconstructed field presents as strong twisted regions wrapped by a boundary 
of quasi-separatrix layers~\citep[QSLs,][]{Titov_2007}, 
which in fact is natural according to the definition of the magnetic flux rope~\citep{rliu_2016}. 
Besides the coherent structure of one flux rope, the field lines of which have similar rotational patterns~\citep{rliu_2016}, 
more complex configurations are found in the ARs, such as double-decker configuration with one flux rope atop the other one~\citep{Kliem_2014}, intertwined double-decker flux rope~\citep{Mitra_2020}, 
braided flux ropes~\citep{Awasthi_2018}, co-existent flux rope and sheared arcade in one filament~\citep{Guo_2010}, etc.

Overall,  
$\delta$ sunspot which indicates a non-potential configuration, and flux rope which is a highly non-potential structure in the corona, both play important role in solar eruptions. 
Detailed study of the magnetic flux rope in $\delta$ sunspot region, including its formation, configuration and eruption, and moreover, the role that $\delta$ spot configuration plays in above process, is necessary in understanding solar eruptions. 
A few observations do have related the two explicitly. 
~\citet{Jiang_2012b} reported that a highly twisted filament channel was formed above the $\delta$ sunspot in NOAA AR 11158. 
With the aid of coronal field extrapolation,~\citet{Mitra_2018a} also found that the flux ropes successively erupted from AR 12673 were formed in the $\delta$ sunspot of the AR. 
In this work, we investigate the configuration and failed eruption of a complex flux rope structure above the central $\delta$ spot of AR 11515. The AR is very prolific in flares and CMEs and has been intensively studied~\citep{Louis_2013, Jingj_2014, Wang_2014, Wang_2018e}. 
The flux rope we studied is related to a confined M5.3-class flare occurred on 2012 July 4. 
The paper is structured as follows: 
in  the next section, we describe the data and method used. The results are presented in Section~\ref{sec:res}. We discuss the implication of this work in Section~\ref{sec:dis}. 

\section{Data Analysis}\label{sec:dat} 

We use the photospheric continuum intensity maps and vector magnetograms provided by the Helioseismic and Magnetic Imager~\citep[HMI,][]{Scherrer_2012} on board the {\it Solar Dynamics Observatory)}~\citep[{\it SDO},][]{Pesnell_2012} to investigate the evolution of the $\delta$ sunspot. 
HMI generates the vector magnetograms with a plate scale of 0\farcs 5 and a cadence of 12 minutes, 
based on which the ARs are automatically extracted and tracked. 
Here the data segment deprojected to cylindrical equal area (CEA) coordinate is used~\citep{Bobra_2014}. The data segment is released as HMI.sharp$\_$cea$\_$720s series~\citep{Hoeksema_etal_2014}.  
Using the field data, we calculate the unsigned magnetic flux via $\Phi=\int_s |B_z|ds$, in which $B_z$ is the vertical component of the magnetic field, and $ds$ is the area of the pixel. Since the magnetic field measurement suffers from severe projection effect when near the solar limb~\citep{Petrie_2015}, we only calculate $\Phi$ when the AR is not far from the disk center, i.e., in the region between the Stonyhurst longitude $60^{\circ}$E to $60^{\circ}$W. 

We employ the ultraviolet (UV) and extreme-ultraviolet (EUV) images provided by the Atmospheric Imaging Assembly~\citep[AIA,][]{Lemen_etal_2012} onboard {\it SDO} to inspect the 
details of the M5.3-class flare. 
The data has a plate scale of 0\farcs 6 and a cadence up to 12s. 
The hard X-ray (HXR) emission of the flare is recorded by the {\it Reuven Ramaty High-Energy Solar Spectroscopic Imager}~\citep[{\it RHESSI},][]{Lin_2002}.  
We use the detectors 1F, 3F, 4F, 5F, 6F, 7F and 8F to reconstruct the HXR sources at 12-25 KeV by the Clean algorithm~\citep{Hurford_2002}.

We study the coronal magnetic field of the region by extrapolating the photospheric vector magnetogram using a non-linear force free field (NLFFF) method~\citep{Wiegelmann_2004, Wiegelmann_2012}. 
Since the photosphere is not necessarily force-free, the photospheric magnetogram used as the input of the extrapolation needs to be preprocessed to reduce possible force, torque, and other noises~\citep{Wiegelmann_2006}. Three dimensionless parameters, the flux balance parameter, the force balance parameter, and the torque balance parameter, are calculated to quantify the quality of the 
preprocessed magnetogram. 
We also calculate the fractional flux ratio and the weighted angle between current ($\bf J$) and magnetic field ($\bf B$) to check the divergence-freeness and force-freeness of the extrapolated field~\citep{Derosa_2015}.  

Based on the extrapolated field, 
we calculate the twist number $T_w$ and squashing factor $Q$ 
using the code\footnote{\url{http://staff.ustc.edu.cn/~rliu/qfactor.html}} developed by~\citet{rliu_2016} to identify the flux rope, the cross section of which is manifested as areas of high $T_w$  wrapped by high $Q$ boundaries~\citep{rliu_2016}. $T_w$ measures the number of turns that two extremely close field lines wind about each other, and is computed through equation $\displaystyle T_w = \frac{1}{4\pi}\int_l\alpha dl$~\citep{Berger_2006a, rliu_2016}, in which $\alpha$ denotes the force free parameter and $dl$ denotes the elementary length of the field line. $Q$ quantifies the local gradient of the magnetic connectivity. For an arbitrary field line, its two footpoints define a mapping: $\bm{r_1}(x_1,y_1) \rightarrow \bm{r_2}(x_2,y_2)$, the Jacobian matrix of which is $\displaystyle D_{12}=\bigg[\frac{\partial \bm{r_2}}{\partial \bm{r_1}}\bigg]=\Bigg( \begin{matrix}
\frac{\partial x_2}{\partial x_1} & \frac{\partial x_2}{\partial y_1}\\
\frac{\partial y_2}{\partial x_1} & \frac{\partial y_2}{\partial y_1}
\end{matrix} \Bigg) \equiv \Bigg( \begin{matrix}
a & b\\
c& d
\end{matrix} \Bigg)$. The squashing factor is then defined as $\displaystyle Q \equiv \frac{a^2+b^2+c^2+d^2}{|B_{n,1}(x_1,y_1)/B_{n,2}(x_2,y_2)|}$, in which $B_{n,1}(x_1,y_1)$ and $B_{n,2}(x_2,y_2)$ are components perpendicular to the plane of the footpoints~\citep{Titov_2002, rliu_2016}. High $Q$ regions usually indicate the QSLs, 
a thin volume where the connectivity of magnetic field lines has drastic change. 
The electric current is easily to be accumulated at the QSLs, favoring of magnetic reconnection. 
For a coherent flux rope possessing some degree of cylindrical symmetry, we can further take the field line owning the extremum $T_w$ of the flux rope as a reliable proxy of its axis~\citep{rliu_2016}.  
We also search the null point through solving the equation $B_i(x,y,z)=0$, in which $i=x,y,z$, using a modified Powell hybrid method\footnote{\url{https://lesia.obspm.fr/fromage/}}~\citep{Demoulin_1994}. 
We then visualize the NLFFF field lines via a software named as PARAVIEW\footnote{\url{https://www.paraview.org/}}, which provides an interactive environment for visualization. 

\section{Results}\label{sec:res}

\subsection{Formation of the $\delta$ sunspot}\label{subsec:spot}

\begin{figure*}
\begin{center}
\includegraphics[width=1\hsize]{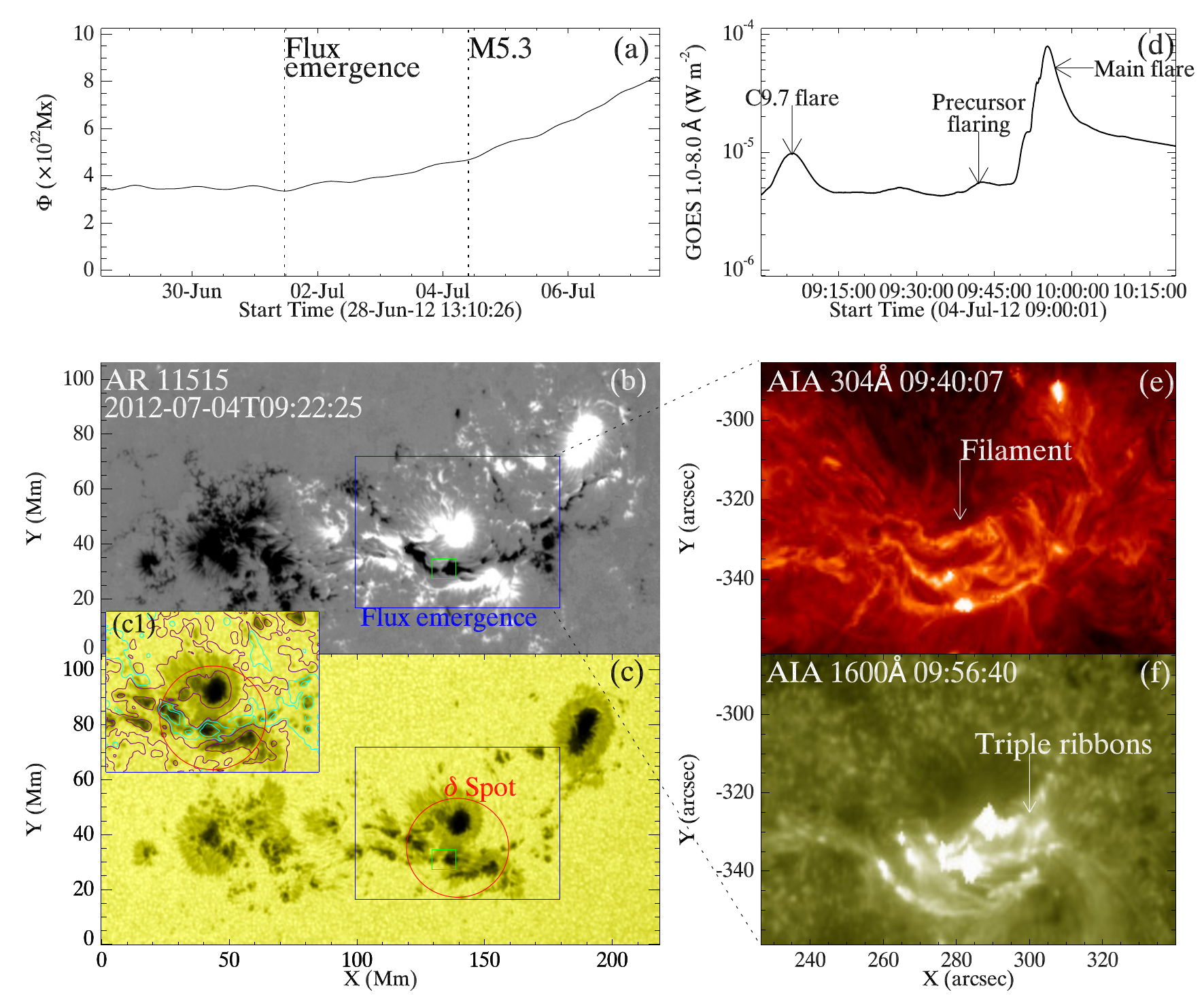}
\caption{Overview of the AR and the flare. 
(a) Temporal evolution of the unsigned magnetic flux $\Phi$ of the AR. $\Phi$ is calculated from 2012 June 28 13:00 to 2012 July 7 11:10, i.e., when the center of the AR has Stonyhurst longitude less than $60^{\circ}$.  
The two vertical-dashed lines mark the onset instants of the flux emergence and the flare. 
(b) $B_z$ component of the photospheric vector magnetic field of the AR at 09:22 on 2012 Jul 4, 
with white and black colors indicating the positive and negative polarities, respective, saturating at $\pm 1200$ G. 
(c) Continuum image of the AR. 
The blue rectangle in (b) and (c) marks the field of view (FOV) of panels (e), (f) and Figure~\ref{fig:bz_cont}. 
The green rectangle marks the FOV of Figure~\ref{fig:bp_config}. 
The continuum intensity in the blue   region is further shown in the inset (c1), with contours of $B_z$ overplotted. The contour levels are -1000, -100, 100 and 1000 Gauss, with positive (negative) values shown in purple (cyan).  
The red circle in (c) indicates the $\delta$ sunspot region. 
(d) GOES flux of the M5.3 class flare. The three arrows indicate a small C9.7-class flare, 
and the precursor and main phase of the M5.3-class flare. See details in Section~\ref{sec:res} and Section~\ref{sec:dis}.  
(e) One snapshot in the AIA 304~\AA~passband prior to 
the flare, showing the filament. 
(f) One snapshot in the AIA 1600~\AA~passband, showing multiple ribbons during 
the flare. 
}\label{fig:overview}
\end{center}
\end{figure*}

When appears at the east limb on 2012 June 27, 
NOAA AR 11515 is already developed with a complex $\beta\gamma$ configuration. 
It has unsigned magnetic flux as large as 3.5$\times10^{22}$~Mx when passing the Stonyhurst longitude $60^{\circ}$E on 2012 June 28 (see Figure~\ref{fig:overview}(a)). 
About 3 days later, a new episode of flux emergence initiates in the western part of the AR, which transports more than 4.5$\times10^{22}$~Mx to the AR 
before it rotates out of the view. 
During the emergence, a $\delta$ sunspot is formed in the middle of the AR (Figure~\ref{fig:overview}(b)-(c)). 
An M5.3 class flare, accompanied by the failed eruption of a filament, 
takes place in the $\delta$ sunspot region from 09:47 to 09:57 UT
on 2012 July 04 (Figure~\ref{fig:overview}(d)-(f)). 
The reason for the failure of the eruption has been studied in~\citet{Li_2019}. 
Here we mainly focus on the formation of the $\delta$ sunspot, the detailed configuration of the pre-eruption flux rope and the magnetic topology accounting for the eruption characteristics. 

\begin{figure*}
\begin{center}
\includegraphics[width=1\hsize]{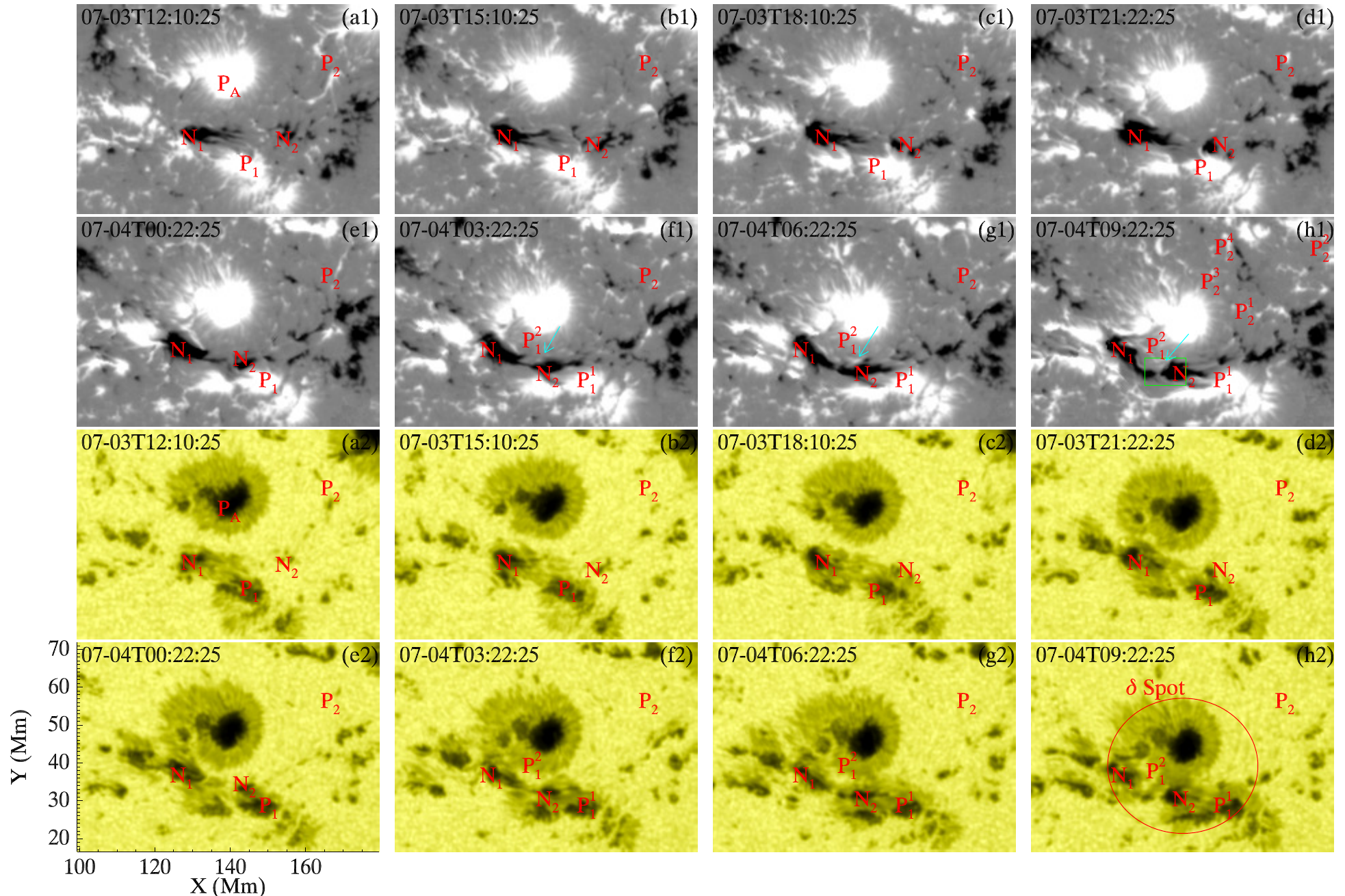}
\caption{Formation of the $\delta$ sunspot. Panels (a1)-(h1) display the photospheric $B_z$ magnetograms, with B$_z$ saturating at $\pm 1200$ G, while panels (a2)-(h2) are the continuum images. Label P$_A$ denotes a pre-existent positive polarity. N$_1$ and P$_1$ indicate the negative and positive polarities of bipole1, respective, while N$_2$ and P$_2$ are for bipole2. P$_1^1$ and P$_1^2$ denote two patches of P$_1$. 
The newly formed P$_1^2$, which is located at the north of N$_2$, is further indicated by a cyan arrow in panels (f1)-(h1).  
The labels P$_2^1$, P$_2^2$, P$_2^3$ and P$_2^4$ in panel (h1) are discrete patches of P2, corresponding to different field lines in Figure~\ref{fig:rope_layers}. The green rectangle in panel (h1) indicates the FOV of Figure~\ref{fig:bp_config}. The red circle in (h2) marks the $\delta$ sunspot configuration. An associated animation is available online. 
}\label{fig:bz_cont}
\end{center}
\end{figure*}

While the flux emergence starts from 2012 July 01, the $\delta$ sunspot is not formed until 2012 July 04. 
We present the emergence period relevant to the spot formation in Figure~\ref{fig:bz_cont}. 
On July 03, 
a newly emerged bipole, bipole1, 
has drifted from the west to the middle of the AR (Figure~\ref{fig:bz_cont}(a1)),  
followed by another small bipole, bipole2. 
The two polarities of bipole1 (N1 and P1) are observed 
as developed sunspots consisting of the umbra and penumbra in the continuum image (Figure~\ref{fig:bz_cont}(a2)), while the two polarities of bipole2 (N$_2$ and P$_2$) are presented as pores.
As emergence goes on, the polarity N$_2$ moves to the east, and finally collides with P$_1$ (Figure~\ref{fig:bz_cont}(b1-e1)).  
After collision, the flux emergence continues. 
Due to the existence of N$_2$, some newly emerged positive flux of bipole1 
forms a small patch that is isolated from the original P$_1$, being located at the north of N$_2$.   
We call the new patch 
as P$_1^2$ (indicated by a cyan arrow in Figure~\ref{fig:bz_cont}(f1-h1)), and the old, large P$_1$ as P$_1^1$ (Figure~\ref{fig:bz_cont} (f1-h1)). P$_1^2$ appears as a patch of penumbra in the continuum image (Figure~\ref{fig:bz_cont}(f2-h2)). 
Simultaneously, N$_2$ also develops into a sunspot in the continuum image, 
which together with P$_1^1$ and P$_1^2$ form the $\delta$ sunspot configuration (enclosed by the red circle in Figure~\ref{fig:bz_cont}(h2)). 
Note that P$_2$ is still dispersed, consisting of small flux patches that appear as small pores in the continuum image.

\subsection{Configuration of the flux rope}\label{subsec:rope}

\begin{figure*}
\begin{center}
\includegraphics[width=1\hsize]{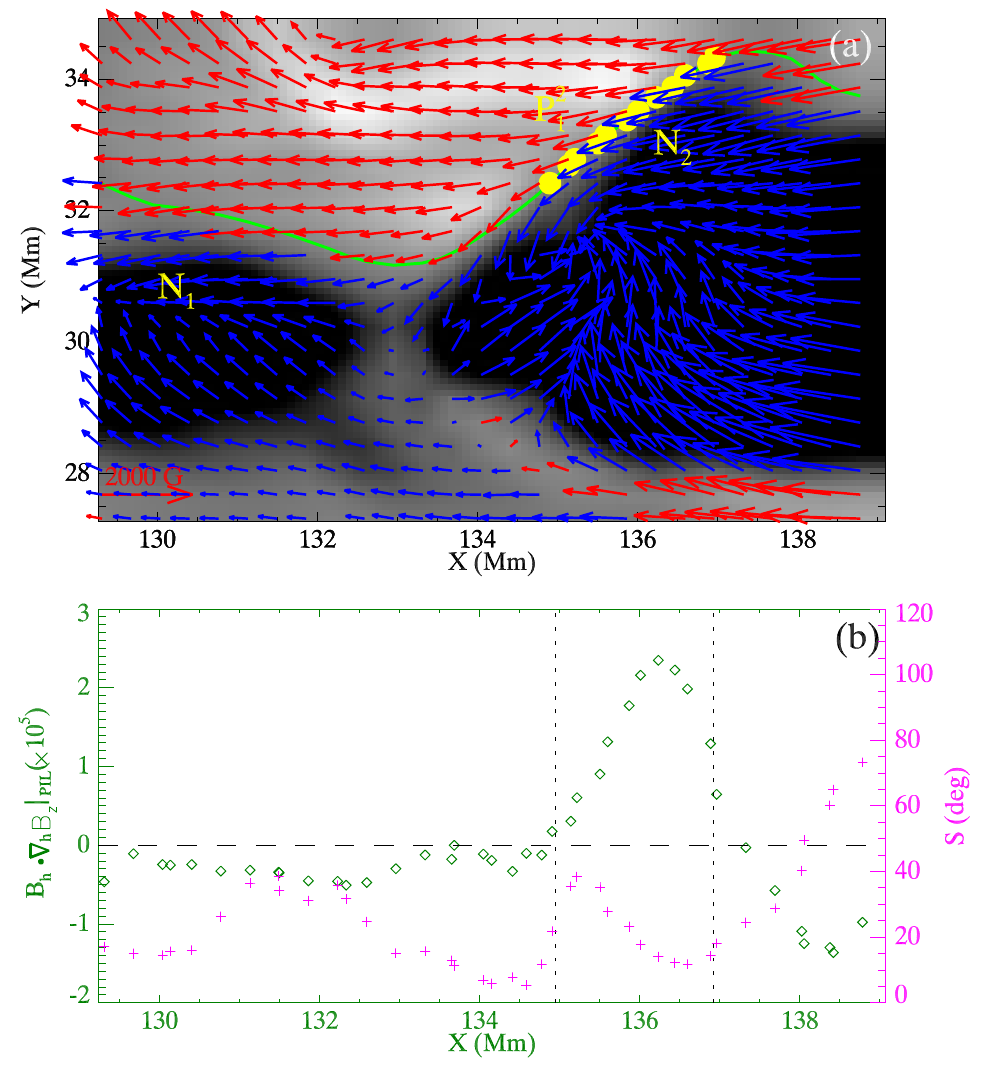}
\caption{The bald patch between N$_2$ and P$_1^2$. 
(a) The photospheric vector magnetogram of the BP region. The FOV of this panel is indicated by the green box in Figure~\ref{fig:bz_cont}(h1). 
The B$_z$ component of the field is plotted as the background, saturating at $\pm 600$ G.  
The red (blue) arrows denote the horizontal component (B$_h$) of the field that are corresponding to the positive (negative) polarities. 
The green line indicates the PIL between P$_1^2$ and the group of N$_1$ and N$_2$, which is obtained at the contour line of $B_z=0$. 
The yellow parts of the PIL mark the BP section. 
(b) Values of $({\bm B_h\cdot\nabla_h B_z})|_{PIL}$ (shown in green) and the shear angle $S$ (shown in magenta) of all pixels along the PIL. 
The two vertical-dashed lines in (b) enclose the BP region. 
}\label{fig:bp_config}
\end{center}
\end{figure*}

We further select a moment (2012-07-04T09:22 UT) to investigate the pre-flare magnetic condition above the $\delta$ sunspot region. 
The photospheric vector magnetic field and the extrapolated coronal field are checked, shown in Figure~\ref{fig:bp_config} to Figure~\ref{fig:rope_layers}. 
We find a sheared PIL formed between P$_1^2$ and the group of N$_1$ and N$_2$ (green line in Figure~\ref{fig:bp_config}(a)), %
which is obtained via drawing the contour of $B_z=0$ on the $B_z$ map directly. 
The pixels of the PIL all have shear angle ($S$) larger than 55$^{\circ}$ (Figure~\ref{fig:bp_config}(b)), indicating the high non-potentiality here. 
$S$ is the angle between observed field (${\bm B_{Obs}}$) and the potential field (${\bm B_{Pot}}$), calculated by 
 $S=\displaystyle cos^{-1}(\frac{{\bm B_{Obs}}\cdot\bm B_{Pot}}{|B_{Obs}||B_{Pot}|})$~\citep{Bobra_2014}. 
The potential field is computed by a frourier transformation method~\citep{Alissandrakis_1981}. 
Moreover, we find a bald patch (BP) formed at the interface between N$_2$ and P$_1^2$ (yellow dots in Figure~\ref{fig:bp_config} (a)), i.e., between one umbra and the penumbra of the $\delta$ spot. 
BP is a configuration at which the field lines are tangent to the photosphere and concaved up, showing a inverse configuration that points from the negative polarity to the positive polarity (see blue vectors in Figure~\ref{fig:bp_config} (a)). 
It can be identified through the discriminant $({\bm B_h\cdot\nabla_h B_z})|_{PIL}>0$~\citep[][see Figure~\ref{fig:bp_config}(b) here]{Titov_1993}, where $B_h$ and $B_z$ denote the horizontal and vertical components of the magnetic field. 
BP usually belongs to the bald patch separatrix surface (BPSS), 
which is prone to the formation of a flux rope. 
Such configuration is found to be preferred location for current accumulation in the corona and thus magnetic reconnection~\citep{Titov_1999}. 

We then search for the flux rope in the extrapolated NLFFF. The quality of the preprocessed photospheric magnetogram is firstly checked. The values of the dimensionless parameters (see Section~\ref{sec:dat}) are presented in Table~\ref{tb:para}. It is found that the value of the flux balance parameter slightly increases after preprocessing, but is still of the order of $10^{-2}$, indicating that the magnetic flux of the magnetogram is basically balanced, either before or after the preprocessing. The values of the force balance and torque balance parameters are of the order of $10^{-2}$ after preprocessing, decreasing by an order of magnitude compared to the values before preprocessing, 
suggesting that the force and torque on the magnetogram 
are significantly reduced. 
We also check the quality of the extrapolation. It is found that the value of the fractional flux ratio is $2.03\times10^{-3}$, and the weighted angle between $\bf J$ and $\bf B$ is $9.07^{\circ}$. 
While the former value is small, the latter is about $2^{\circ}$ larger than the values of some reported NLFFF results~\citep{Lliu_2017, Mitra_2020}. 
Nevertheless, both values are not large, suggesting that the degree of the divergence-freeness and force-freeness of the extrapolation is acceptable.  

We do find a flux rope structure (shown in Figure~\ref{fig:rope}(a)-(b)) above the $\delta$ spot region 
through checking the $T_w$ and $Q$ distributions in a series of vertical planes across the PIL. 
The flux rope displays a complex configuration consisting of different branches, %
with one compact end rooted in N$_1$, and the other end bifurcating into different branches that are rooted in P$_1^1$, 
discrete flux patches of P$_2$, and also part of the periphery of the pre-existent P$_A$. 
In the following, we call the 
branch rooted in P$_1^1$ as the southern branch of the flux rope (indicated by a pink rectangle in Figure~\ref{fig:rope}(a)), since P$_1^1$ is located in the southern part of the AR, and the ones rooted in P$_2$ and P$_A$ (enclosed in a yellow rectangle) as northern branches. 
The bottom of the flux rope touches the photosphere at the BP between N$_2$ and P$_1^2$ (see the field lines passing through the BP in Figure~\ref{fig:rope}(b1)). 
We further try to identify the axis of the flux rope through inspecting the local extremums of the $T_w$ maps. A field line possessing the local maximum in all $T_w$ maps we inspected is found, which may be deemed as the proxy of the flux rope axis~\citep{rliu_2016}.  
$Q$, $T_w$, and the in-plane field vectors in a vertical plane normal to the tangent line of the apex (local peak of a curve) of the possible axis proxy are displayed in Figure~\ref{fig:rope}(c)-(e). 
Unsurprisingly, the cross section of the flux rope is manifested as the strong twisted region wrapped by high Q boundary. 
Moreover, it exhibits more complex, onion-like configuration, consisting of different layers that are separated by different high Q boundaries. 
The most of the in plane field vectors are found rotating around the possible axis proxy (denoted by the red triangles in Figure~\ref{fig:rope}), excepting the ones below the possible axis. This may be because that the flux rope owns 
bifurcated ends, which indicates that it is not strictly coherent, 
so that there may exist a few field lines which have slightly different rotational patterns from the other field lines at some locations. 
We also compare the flux rope and the filament observed in AIA 304~\AA~passband (Figure~\ref{fig:rope}(f)). It is found that the eastern part of the flux rope coincides with the filament to a large extent. The western part of the filament is covered by higher loops, which makes the direct comparison 
unachievable. 
Moreover, we find two sets of loops in the AIA 131~\AA~passband, which overlie the southern and northern branches of the flux rope, respectively (Figure~\ref{fig:rope}(g)). 

We further check the detailed properties of each layer of the flux rope (Figure~\ref{fig:rope_layers}). Typical field lines of each layer are randomly chosen to show. 
It is found that $T_w$ decreases gradually from the core to the boundary. 
The core layer (layer 5), where the possible axis proxy threads, 
owns the largest twist 
with almost uniform $T_w$ as high as $-3.8$ (Figure~\ref{fig:rope_layers}(b)). 
The field lines of the layer are rooted in the polarities N$_1$ and P$_2^4$ (see the blue line in Figure~\ref{fig:rope_layers}). 
The outer layer, layer 4, has lower $T_w$ roughly ranging from $-3$ to $-2.7$. 
The field lines of this layer originate from the polarity P$_2^3$, 
then go into N$_1$ (see the green line in Figure~\ref{fig:rope_layers}). 
Some of the field lines further pass though P$_1^2$ and N$_2$ (the BP).  
Layer 3 has $T_w$ ranging from $-3$ to $-1.5$. 
Its field lines span the longest distance, rooted in polarities N$_1$ and remote P$_2^2$ (see the orange line in Figure~\ref{fig:rope_layers}).  
Layer 2 owns $T_w$ in the range of $-2.4$ to $-1$. 
Its field lines are found rooted in N$_1$ and P$_2^1$ (see the pink line in Figure~\ref{fig:rope_layers}). Some of the field lines also pass through the BP. 
The outermost layer (layer 1) is the least twisted part, having $T_w$ ranging from $-0.9$ to $-0.5$ (see the red line in Figure~\ref{fig:rope_layers}).   
Its field lines are rooted in polarities N$_1$ and P$_1^1$, with some of them also passing through the BP. 
They indeed do not meet the criterion of the field lines belonging to 
a typical flux rope, which should wind at least one full turn.  
However, they are intertwined with the more twisted inner layers, 
together forming an inseparable whole structure. 
We thus call this structure a complex flux rope.

\subsection{Topological origin of the eruption characteristics}\label{subsec:eruption} 



The flare is accompanied by the failed eruption of a filament. Taking the filament as a proxy of the flux rope, its failed eruption suggests that the flux rope experiences an eruptive expansion during the flare. 
The Lorentz force during the eruption is non-zero, making the in-eruption NLFFF extrapolation unreliable. 
We thus check the eruption characteristics observed in various passbands to deduce the eruptive 
details of the flux rope. We select three AIA passbands, the hot 131~\AA~($\sim$10~MK), cool 304~\AA~($\sim$50000 K), and the UV 1600~\AA~($\sim$10000 K) to show the 
eruptive details (Figure~\ref{fig:eru}). 
The GOES soft X-ray (SXR) flux shows that the flare starts from 07-04T09:47 UT, peaks at 09:55, and ends at 09:57 (Figure~\ref{fig:overview}(d)). 
Besides, 
from around 09:35 to 09:47, a small ``bump'' appears on the SXR curve (Figure~\ref{fig:overview}(d)), suggesting that some mild flarings, 
i.e., a phase of mild reconnection occurs prior to the recorded flare. 
This kind of gradual buildup of the SXR flux prior to the main flare, 
which is distinct from the rapid increasing of the SXR flux during the main flare, is called a precursor phase 
\citep{Zhang_2006c, Mitra_2019}. 
Furtherly, in the hot 131~\AA~passband (Figure~\ref{fig:eru}(a1)-(b1)), an inclined ``v''-shaped structure brightens. 
The structure seems to consist of two sets of loops, with two far sides rooted in P$_1^1$ and dispersed P$_2$, 
and the other two close sides somehow intersecting at higher altitude 
to form the cusp of the ``v'' (Figure~\ref{fig:eru} (a1)). 
In the {\it RHESSI} observation, an HXR source at the 12-25 keV energy bands is found to be centered at the cusp. 
The EUV brightening and HXR source further support that there is a phase of 
precursor reconnection 
occurring prior to the main flare. 
We call this phase of reconnection as reconnection1. 

After the precursor brightening,  
a filament bundle starts to rise slowly from around 09:46 
(see Figure~\ref{fig:eru}(c2)-(d2) and associated movie). 
The footpoints of the filament brighten during the rise, forming an HXR source at the eastern footpoint (Figure~\ref{fig:eru}(d2)). From around 09:52, another brightening occurs below the rising filament, which is visible in both 131~\AA~and 304~\AA~passbands, as well as in the 1600~\AA~passband, forming a new HXR source (Figure~\ref{fig:eru}(e1)). This is the main phase reconnection generating the peak SXR flux of the flare.  
We call this phase of reconnection as reconnection2. 
Following reconnection2, the filament continues to rise and finally stops without forming a CME. 
After the failed eruption, two sets of nearly-potential post-flare loops appear above the $\delta$ spot region, with one set connecting P$_2$ (also part of the periphery of P$_A$) and the group of N$_1$ and N$_2$ (enclosed in a yellow rectangle in Figure~\ref{fig:eru}(f1)), and the other set connecting P$_1^1$ and the group of N$_1$ and N$_2$ (enclosed in a pink rectangle in Figure~\ref{fig:eru}(f1)). 
Three elongated ribbons appear at the footpoints of the loops in the 1600~\AA~passband (Figure~\ref{fig:eru}(f3)). Besides, there is a remote ribbon brightening during the flare, located at the east of the AR, which seems to be connected with the core region by some sheared loops (Figure~\ref{fig:eru}(d1) and (d3)).

\begin{figure*}
\begin{center}
\includegraphics[width=1\hsize]{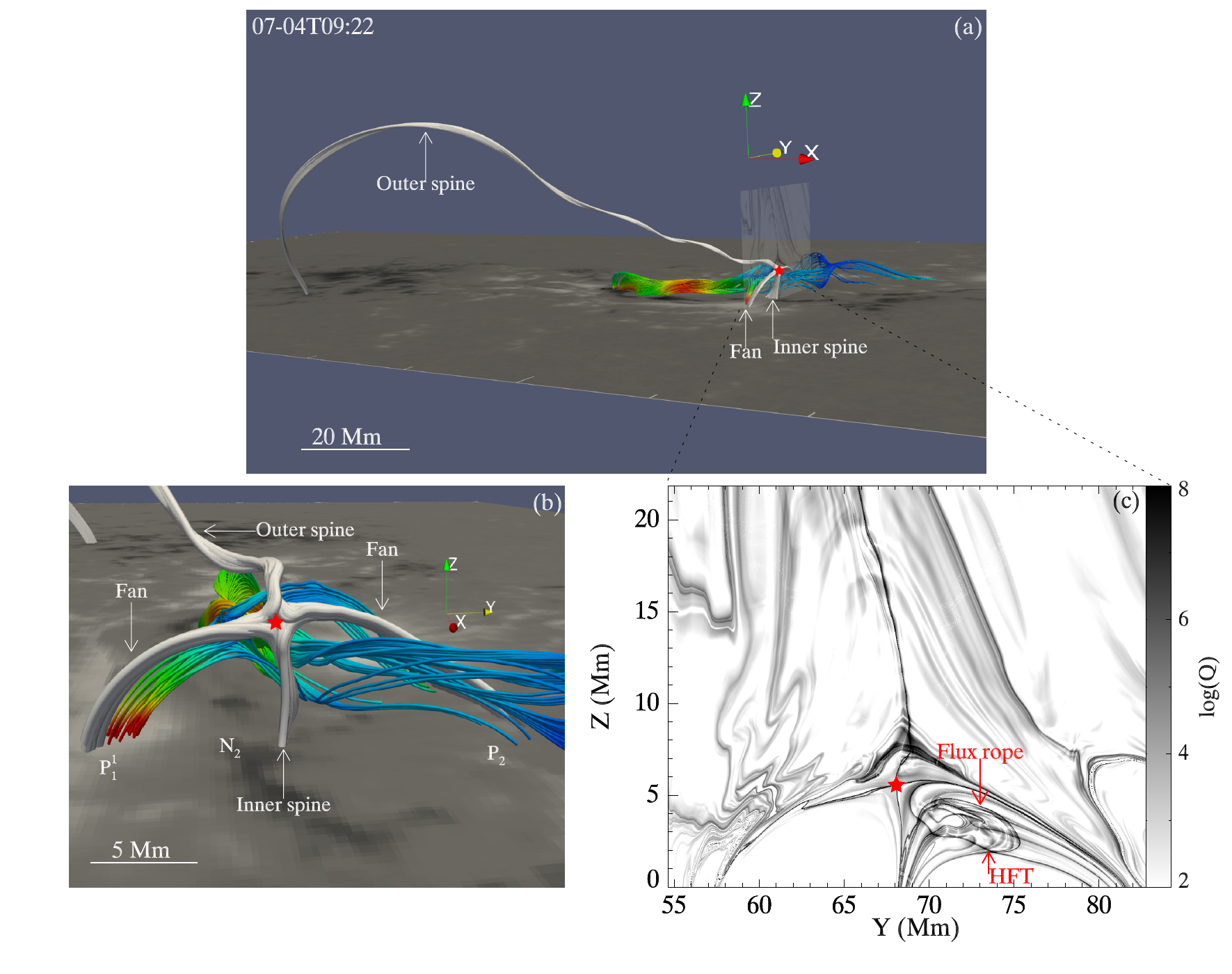}
\caption{The null point above the flux rope, which is marked as a red star. (a)-(b) Different view of the null point. Magnetic field lines traced from different topological structures are shown in different colors. The iridescent lines belong to the flux rope while the grey lines are for the coronal configurations associated with the null point. (c) The distribution of logarithmic $Q$ in a vertical plane crossing the magnetic null. The position of the plane is indicated by a semi-transparent cut in panel (a). The photospheric B$_z$ in panels (a) and (b) saturates at $\pm 2000$ G. 
}\label{fig:null}
\end{center}
\end{figure*}

The inclined ``v''-shaped structure, which shows an HXR source centered at its cusp during the flare, strongly suggests that there might be a null point. 
We thus search the pre-eruption NLFFFs using the method described in Section~\ref{sec:dat}, and do find a null point above the flux rope (Figure~\ref{fig:null}). 
The configuration of the null point is very similar to the classical null associated with 
fans and spines~\citep{Priest_1996a}. 
The outer spine here 
is rooted in the remote negative polarity, while the inner spine is rooted in N$_2$. 
The only exception here is that the field lines of the fans 
are rooted in two disconnected polarity patches, P$_1^1$ and P$_2$, different from the circular fan of the classical configuration. 
The null is manifested as the intersection of two high $Q$ lines (indicated by a red star in Figure~\ref{fig:null}(c)) in a vertical plane across it. 
Moreover, the flux rope is exhibited as a inverse teardrop-shaped structure in this plane, similar as in Figure~\ref{fig:rope}, but having an HFT 
configuration at its bottom. HFT is a configuration where two QSLs intersect, prone to current accumulation, thus reconnection~\citep{Titov_2002a}.

\begin{figure*}
\begin{center}
\includegraphics[width=1\hsize]{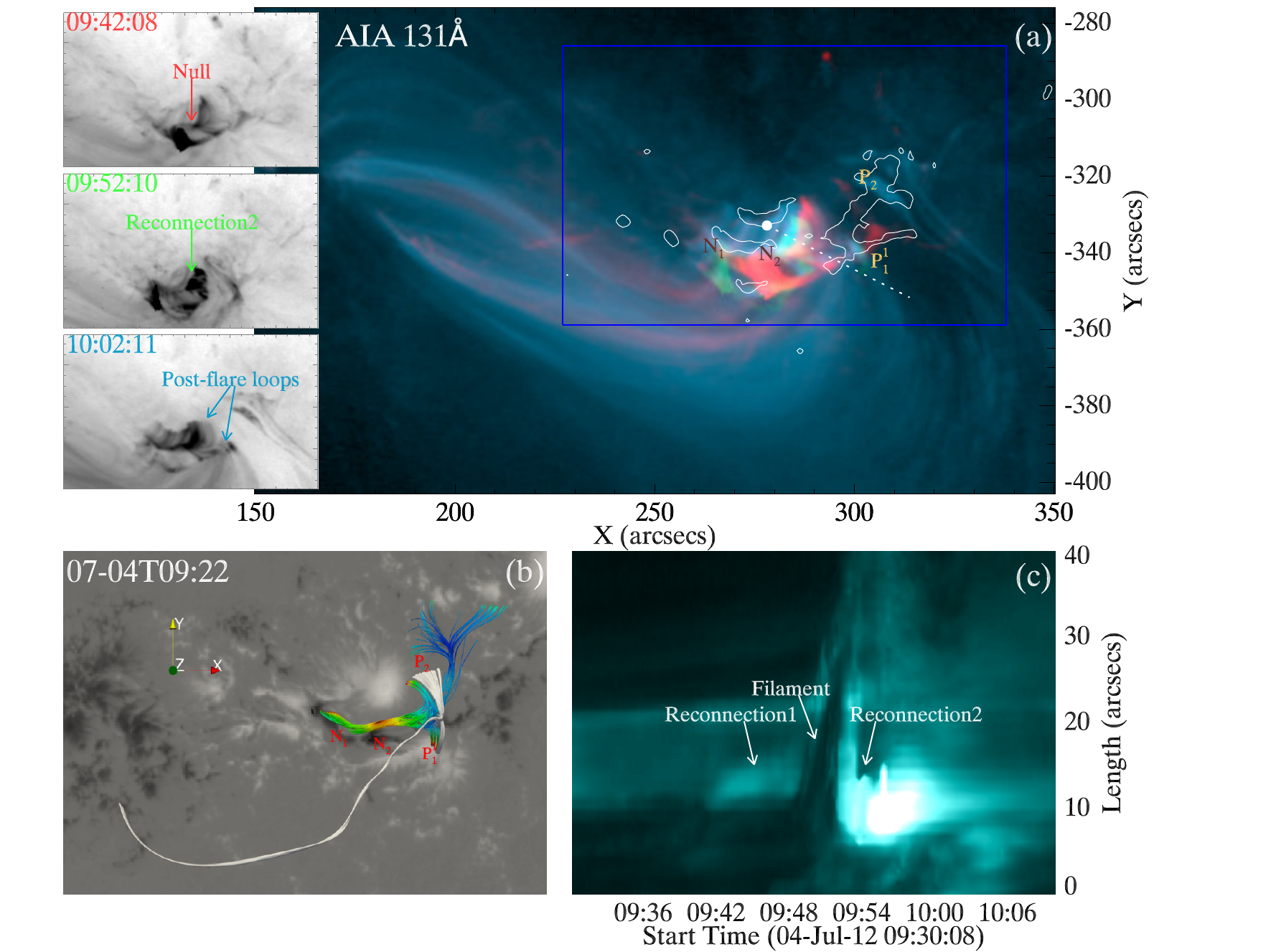}
\caption{Comparison between the eruptive characteristics and the magnetic topology. 
(a) Composite image of AIA 131~\AA~observations at three different times. The observation at 09:42 is shown in red, the one at 09:52 is shown in green, and the one at 10:02 is shown in blue, which display the coronal null, 
the main phase of reconnection, and the post-flare loops, respective. The three insets at the left are the negative images at the three times. The white contours outline the flare ribbons in the 1600~\AA~observation. The blue rectangle has the same FOV as the one in Figure~\ref{fig:overview}(b). The white-dashed line indicates the position of the slice used in panel (c). The white dot marks the start of the slice. 
(b) The flux rope (shown in iridescence) and  the magnetic field lines of the configurations associated with the null (shown in grey) in the extrapolated NLFFF. Panel (b) has roughly the same FOV as panel (a). 
(c) The time-distance plot of the slice, showing the two phase of reconnection and the rise of the filament. 
}\label{fig:cmpo}
\end{center}
\end{figure*}

We further show the AIA observations and the magnetic topological structures together in Figure~\ref{fig:cmpo} to better understand the eruptive 
process. 
The ``v''-shaped structure in the 131~\AA~image (shown in red in Figure~\ref{fig:cmpo}(a)), i.e., the location where the first phase of the reconnection occurs, generally coincides with 
the null point identified in the reconstructed coronal NLFFFs (Figure~\ref{fig:cmpo}(b)).  
The only difference is that the sheared loops connecting the core region and the remote region in the 131~\AA~passband deviate from the outer spine in the NLFFFs. 
The former are rooted in the more northern part of the remote negative polarity, while the latter is rooted in more southern region. 
This may be resulted from the limitations of the NLFFF model, and the projection effect since the NLFFF image is in the CEA coordinate while the AIA images are in the CCD coordinate. 
The main phase reconnection seems to occur below the null point (show in green in Figure~\ref{fig:cmpo}(a)). 
The position of the two sets of post-flare loops (shown in blue in Figure~\ref{fig:cmpo}(a)) and the three main flare ribbons (white contours in Figure~\ref{fig:cmpo}(a)) correspond well with the bifurcated branches 
of the pre-eruption flux rope. 
Specifically, 
the northern set of the post-flare loops connects P$_2$ (also part of the periphery of P$_A$) and the group of N$_1$ and N$_2$, while 
the northern branches of the flux rope 
are also rooted in P$_2$ and N$_1$, passing through N$_2$ at the BP (Figure~\ref{fig:cmpo}(b)). 
The southern set of the loops connects P$_1^1$ and the group of N$_1$ and N$_2$, while the southern branch of the flux rope 
is also rooted in P$_1^1$ and N$_1$ (also passing through N$_2$). The three ribbons are located in P$_2$ (also part of the periphery of P$_A$),  the group of N$_1$ and N$_2$, and P$_1^1$ as well. 
These suggest that the bifurcated flux rope is involved in the failed eruption. The ejective expansion of different branches of the flux rope stretches different sets of strapping field (see Figure~\ref{fig:rope}(g)), forming two sets of post-flare loops and three flare ribbons which are rooted near the footpoints of the flux rope branches.  
The results support that the extrapolated bifurcated flux rope generally coincides with the observation. 

We also present the time-distance diagram of a slice (white dotted line in Figure~\ref{fig:cmpo}(a)) extracted from the 131~\AA~images to better see the sequence of the eruption characteristics (Figure~\ref{fig:cmpo}(c)). The slice runs across the filament. 
It is seen that before the filament rising, there is a phase of mild brightenings occurring above the filament, which corresponds to the precursor reconnection at the null. After the start of the filament rising, a second phase of strong flarings occurs below the filament, which apparently corresponds to the main phase reconnection. 
The diagram confirms the sequence of the eruption characteristics, i.e., reconnection1 occurs at the null at first, followed by the rise of the filament, and then reconnection2 occurs below the filament, after which the filament continues to rise and finally stops. 
These suggest that the trigger of the failed eruption is more likely to be the breakout reconnection at the null, which weakens the confinement of the strapping field and causes the rise of the filament. 
The main phase reconnection occurring below the filament gives positive feedback to the rising filament. 
Since the main reconnection is visible in various passbands, even in 1600~\AA~which monitors the photosphere and the lower chromosphere,  
it is very possible that it occurs rather low, 
might be associated with the pre-existent BPSS or the HFT below the flux rope.

\section{Summary and Discussion}\label{sec:dis}

In this work, 
via combining the multiple wavelengths observations and the coronal magnetic field extrapolation, we investigate the configuration of a complex flux rope structure in NOAA AR 11515, and its failed eruption during a confined M5.3-class flare. 
It is found that the flux rope is formed right above the $\delta$ spot region, consisting of multiple layers, 
with its cross section exhibiting an onion-like configuration. 
Above the flux rope, a null point is identified. 
The eruption process suggests that the precursor reconnection occurs 
at the null point, which is manifested as a inclined ``v''-shaped structure. 
It weakens the overlying confinement,  
causing the slow rise of the flux rope which is indicated by 
a filament. 
The main phase of reconnection occurs below the flux rope, probably at the BPSS or HFT, facilitating the rise 
of the flux rope. The eruptive expansion of the flux rope finally stops without forming a CME.   

\begin{figure*}
\begin{center}
\includegraphics[width=1\hsize]{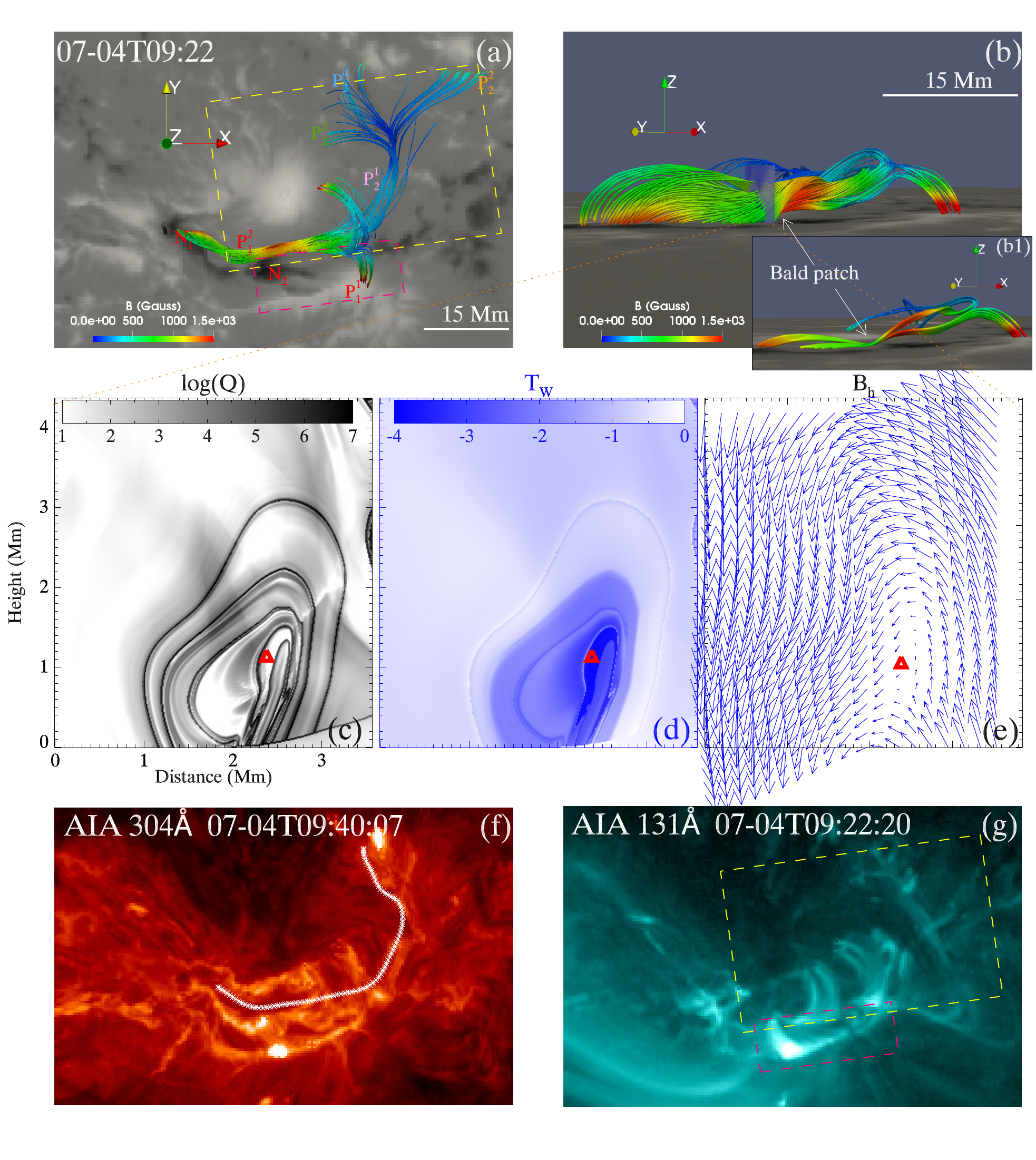}
\caption{The configuration of the flux rope that is identified in the extrapolated NLFFFs. 
(a)-(b) The flux rope seen from two different views. The iridescence colors vary with the strength of the field. Panel (a) has the same FOV as Figure~\ref{fig:bz_cont}. 
The inset (b1) shows the field lines passing through the BP specifically. 
The yellow-dashed rectangle marks the northern branches of the flux rope, while the pink one is for the southern branch. 
The photospheric B$_z$ saturates at $\pm 2000$ G in these panels, with white (black) patches indicating the positive (negative) polarities.  
(c)-(e) The distribution of logarithmic $Q$, $T_w$, and in-plane field vector in a vertical plane which is perpenduclar to the local apex of the possible axis proxy. 
The red triangles indicate the local extremum of $T_w$, i.e., the location where the possible axis proxy 
threads the plane. The position of the plane is indicated by the semi-transparent cut in panel (b).  
(f) The image of the AIA 304~\AA~passband, showing the filament. The white line denotes the estimated axis of the flux rope from NLFFFs. 
(g) The image of the AIA 131~\AA~passband, showing two sets of loops overlying the northern and southern branches of the flux rope. The colored dashed rectangles enclose the same regions 
as the ones in panel (a).  
Panels (a), (f), and (g) have the same FOV. 
}\label{fig:rope}
\end{center}
\end{figure*}

\begin{figure*}
\begin{center}
\includegraphics[width=1\hsize]{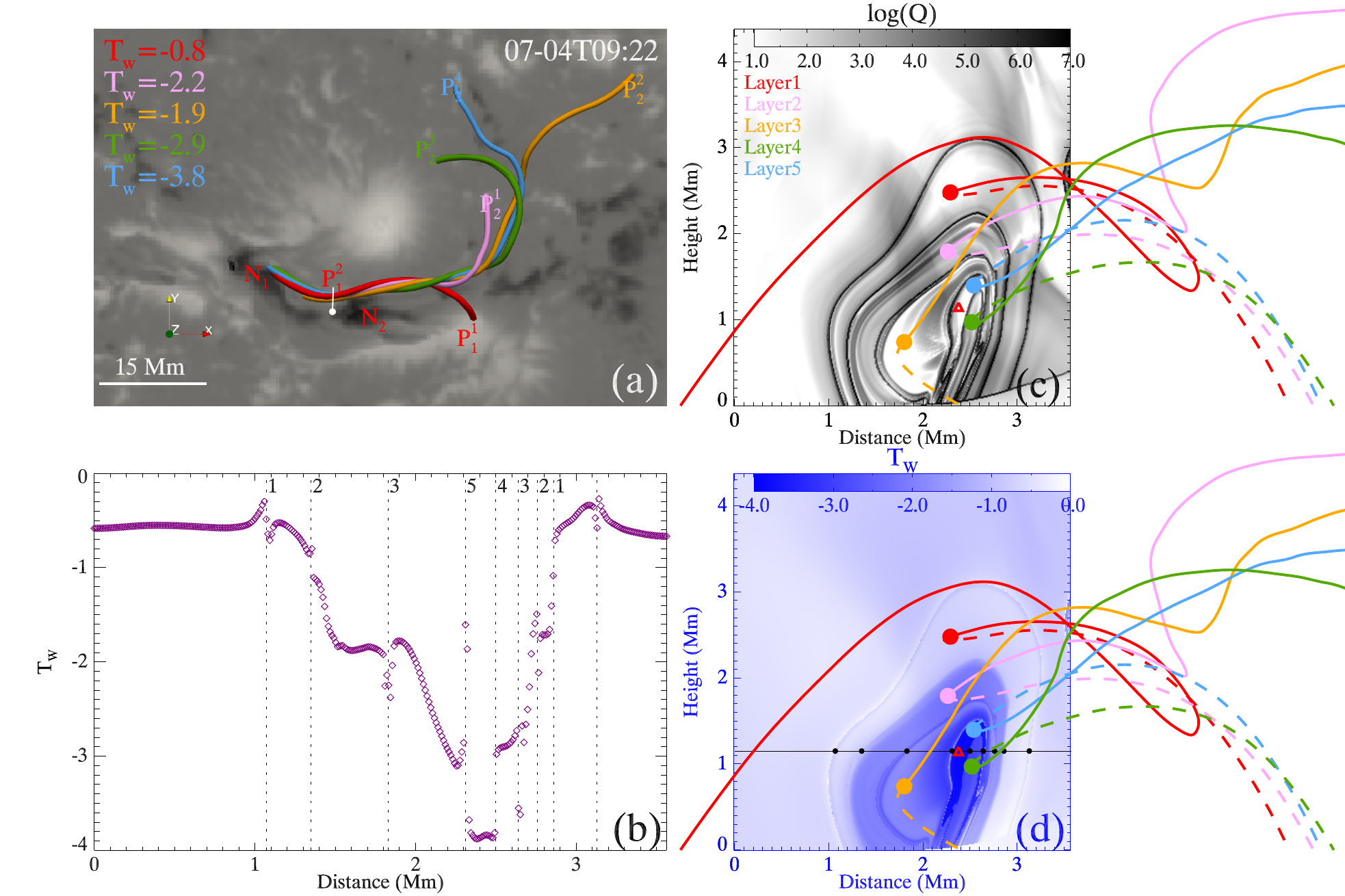}
\caption{Different layers of the flux rope. 
(a) Representative field line from each layer. The panels has the same FOV as Figure~\ref{fig:bz_cont}. The photospheric B$_z$ saturates at $\pm 2000$ G, with white (black) patches indicating the positive (negative) polarities. 
(b) Profile of $T_w$ along the black line in (d), which is parallel to the photopshere and passing the extremum point. 
The vertical-dashed lines correspond to the black dots in panel (d). The digits mark the layers of the flux rope. (c)-(d) Distributions of the logarithmic $Q$ and $T_w$ 
in the same plane indicated in Figure~\ref{fig:rope}(b). The position of the plane is also indicated by a white line in panel (a), with the white dot denoting the start point of the plane.  
The representative field lines from each layer are overplotted. 
The solid parts of the lines are in front of the plane (seen along the line of sight), while the dashed parts are behind the plane. The colord dots mark the positions where the field lines thread the plane. The black dots indicate locations where the horizontal black line intersects with the high Q boundaries. }\label{fig:rope_layers}
\end{center}
\end{figure*}

\begin{figure*}
\begin{center}
\includegraphics[width=1\hsize]{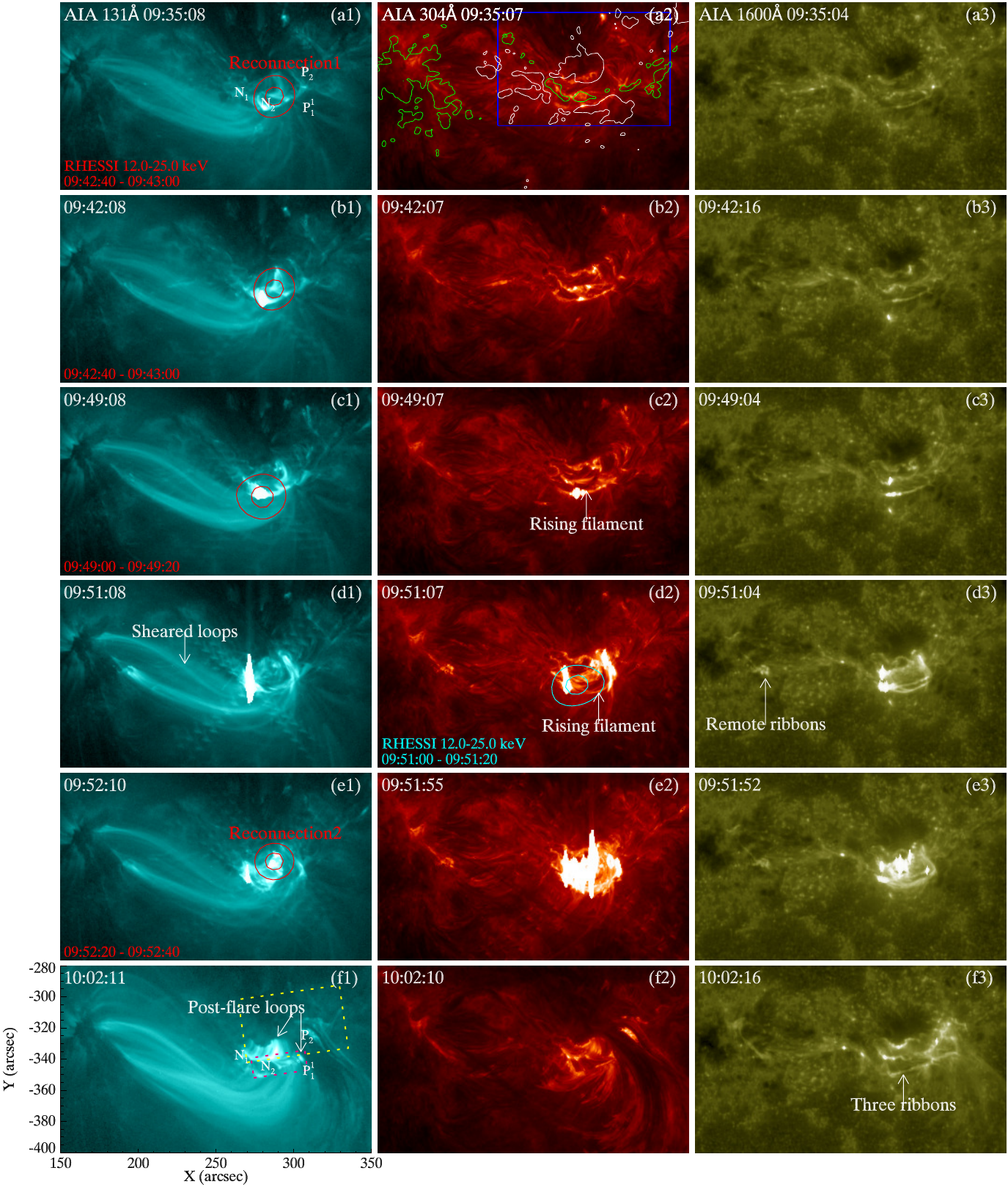}
\caption{Eruptive 
characteristics of the flare. 
(a1)-(f1) Eruptive 
details observed by the AIA 131~\AA~passband. 
The white arrows in (f1) point out two sets of post-flare loops. The two dashed box in (f1) are the same as the ones in Figure~\ref{fig:rope}(g), enclosing the northern and southern sets of loops. 
The polarities P$_1^1$, N$_1, $P$_2$ and N$_2$ are marked in panels (a1) and (f1) for reference. 
(a2)-(f2) Eruption details captured by AIA 304~\AA~passband, showing the eruption of the filament. 
The photospheric $B_z$ is overplotted in panel (a2) for comparison, with white (green) contours for the positive (negative) polarities. The contour levels are -600~G and 600~G. 
(a3)-(f3) Flare ribbons observed in the 1600~\AA~passband. 
The {\it RHESSI} HXR sources in the 12-25 keV energy bands are overplotted in panels (a1), (b1), (c1), (d2) and (e1), with contour levels at $60\%$ and $90\%$ of the peak flux. 
The blue rectangle in panel (a2) has the same FOV as the one in Figure~\ref{fig:overview}(b). An associated animation lasting form 09:00 to 10:10 on 2012 July 4 is available online, which incorporates the details of the precursor and the main flare, and a C9.7-class flare lasting from 09:00 to 09:09.  Brightenings at the BP region and flare ribbons during the C9.7-class flare are indicated by arrows in the movie. The BP region is also enclosed by a green rectangle which is the same as the one in Figure~\ref{fig:overview}(b). See relevant discussion in Section~\ref{sec:dis}.   
}\label{fig:eru}
\end{center}
\end{figure*}

The $\delta$ sunspot is formed through the collision between two spot groups which are newly emerged 
near the pre-existent spot regions (P$_A$).  
The collision occurs between one umbra and the penumbra which become part of the $\delta$ sunspot later. 
A bald patch configuration is found formed at the collision location. 
The identified flux rope is attached to the photosphere 
at the BP (see Figure~\ref{fig:rope}(b1)).   
The result suggests that the $\delta$ sunspot configuration, where two 
umbrae are located close in the same penumbra that contain opposite-signed polarities huddling together, naturally creates an environment prone to the formation of the flux rope. The detailed formation of the flux rope is beyond the scope of this paper. We do not identify flux ropes in extrapolated coronal field before 09:00 on July 4, but do 
find a smaller flare (C9.7-class) that occurs from around 09:00 to 09:09, 
during which small-scale brightenings are observed in the BP region in 1600~\AA, accompanied by brightenings of nearby coronal loops (visible in 131~\AA) which are associated with two other small ribbons 
(see the movie associated with Figure~\ref{fig:eru}). It is suggested that the flux rope may be formed or built-up in confined flares prior to large eruptions~\citep{Guoy_2013, Joshi_2016a, James_2017, Liu_2018d, LLiu_2019}, most probably via magnetic reconnection between different sheared loops in a tether-cutting manner~\citep{Syntelis_2019}.  
Therefore, the flux rope here may be formed through small-scale reconnection between different sheared field lines near the BPSS.  

The flux rope consists of multiple layers, exhibiting an onion-like configuration in the $T_w$ and $Q$ maps of its cross section. 
One of the flux rope end is compact, rooted in one negative polarity, 
while the other end bifurcates into different branches rooted in discrete positive polarity patches, 
indicating that the connectivities of the different branches change drastically to form 
the QSLs around each layer. 
The twist profile of the flux rope is not uniform, with $T_w$ decreasing from the core to the boundary. 
The core has a twist as high as 3.8 turns. 
This is consistent with some in-situ observations of interplanetary magnetic clouds, 
which exhibit highly twisted core enclosed by a less twisted envelope~\citep{Hu_2014, Wang_2018b}. 
The outmost layer of the flux rope here 
has twist less than one full turn, which indeed is ``sheared'' rather than ``twisted''. 
This is similar to the observation reported in~\citet{Guo_2010}, in which a flux rope and sheared arcade sections are found coexisting along one filament. The difference here is that the 
sheared layer is intertwined with the other twisted branches of the flux rope, forming a complex structure. 
Similar but different configurations of complex flux rope have been reported. For example, \citet{Awasthi_2018} identified a complex structure consisting of multiple flux ropes or flux bundles braiding about each other in extrapolated NLFFFs. 
\citet{Hou_2018} also concluded that a multi-flux-rope system is formed through interactions between different emerging dipoles.  
Overall, the observations suggest that the flux rope in $\delta$ sunspot region, or more generally, in multi-polar regions, may have quite complex configuration due to the interaction between adjacent polarities, 
may be different from the simple configuration of a single coherent flux rope in bipolar regions. The latter is often presented in numerical simulations~\citep[e.g.,][]{Gibson_etal_2004, Aulanier_torok_2010}. 


Since the NLFFF assumption does not stand 
during the flare, we compare the pre-eruption magnetic condition and the eruption characteristics observed by various passbands to deduce the details of the eruptive expansion 
of the flux rope.  
The ``v''-shaped structure that brightens prior to the rise of the filament 
corresponds well with the null point identified in the NLFFFs, 
suggesting a precursor phase of reconnection occurs at the null, which is also 
evidenced by the mild enhancement of the SXR flux and the HXR source at the cusp of the ``v''. The precursor may have triggered the flux rope to rise via weakening the confinement of the overlying field, fitting the breakout model. 
This is consistent with previous works, which suggest that the precursor activities, which are associated with small-scale reconnection, have potential to destabilize the flux rope~\citep{Sterling_2005, Sterling_2011, Joshi_2013a, Mitra_2019}. 
Moreover, 
the inner spine and the fans associated with 
the null point are rooted in the polarities forming the $\delta$ spot, again addresses the important role that the $\delta$ spot plays in creating the topology favoring of reconnection and eruption. 

During the main phase of the flare, three parallel ribbons are formed in the polarities where the extrapolated flux rope is rooted, followed by the formation of post-flare loops. The HXR sources are also found at the footpoint of the filament and the looptop. These features are all predicted by the 2D standard flare model (see Section~\ref{sec:intr} for introduction), suggesting that the flare 
fits the standard flare model. 
Moreover, the flaring is visible even in UV observations, suggesting that the reconnection occurs rather low. Considering that there are BP and rather low HFT configurations at the bottom of the flux rope, which are prone to magnetic reconnection, the main phase flare 
is very possible to be associated with those topology. This is not conflicting to the standard flare model 
since in the 3D extension of the standard model~\citep{Aulanier_2012, Janvier_2013}, the elongated current layer surrounding the HFT is thought to be 
an analogy to the current sheet formed between two legs of the strapping field in the 2D model. 

Note that the flare is confined. 
The reason for the failure of the eruption has been investigated by~\citet{Li_2019}. Through calculating the decay index which measures the confinement of the strapping field above the eruption region, 
they concluded that a ``saddle-like'' distribution of the decay index, which is exhibited as a local torus-stale region sandwiched by two torus-unstable regions~\citep{Wangd_2017, Liu_2018a}, is responsible to the failed eruption. The local torus-stable region may have prevented the flux rope from a full eruption. 

The multi-layer configuration of the flux rope may also give hint to the origin of homologous eruptions. 
Homologous eruptions originate from the same magnetic source region, having similar appearance~\citep{Zhang_Wang_2002, Lliu_2017}. It is suggested that those eruptions may be resulted from partial expulsions of the flux rope~\citep{Gibson_2006, Chengx_2018}. 
Although the flux rope studied here seems to expand as a whole, 
we cannot exclude the possibility that in some occasions, 
internal reconnections may occur at the distinct QSLs, resulting in independent eruptions of different layers, thus partial eruptions of the flux rope. 

To summarize, the flux rope we studied is formed right above the $\delta$ spot region, being attached at the interface between one umbra and the penumbra, owning multi-layer configuration with twist decreasing from the core to the boundary. 
The eruptive expansion of the flux rope is triggered by the precursor reconnection at the null point, and facilitated by the main phase reconnection at the BPSS or HFT. The results give details about how complex magnetic topological structures formed in the $\delta$ spot region work together to produce the activities. 

\begin{table}
\begin{center}

\caption{Values of the dimensionless quality parameters of the photospheric magnetogram before and after preprocessing.}\label{tb:para} 
\begin{tabular}{cccc}
\hline \\
Parameter & Before & After \\ 
\hline \\
Flux balance & $1.24\times 10^{-2}$ & $1.41\times 10^{-2}$  \\
Force balance & $0.21$ & $3.07\times 10^{-2}$  \\
Torque balance & $0.18$ & $3.09\times 10^{-2}$ \\
\hline \\ 
\end{tabular} \\
\end{center}
\end{table}

\begin{acknowledgements}
We thank our anonymous referee for the constructive comments which help to improve the manuscript significantly. L.L acknowledges the support from the National Nature Science Foundation of China (NSFC) via grant 11803096, from the Fundamental Research Funds for the Central Universities (19lgpy27), and from the Open Project of CAS Key Laboratory of Geospace Environment. 
JL acknowledges  the support from the Leverhulme Trust via grant RPG-2019-371. 
Y.W. acknowledges the support from NSFC via grants 41574165 and 41774178. 
J.C. acknowledges the support from NSFC via grants 41525015 and 41774186.  
\end{acknowledgements}

\bibliographystyle{aa} 
\bibliography{MFR_11515.bib} %
\end{document}